# Nonlinear Focal Modulation Microscopy

**Authors**

Guangyuan Zhao[1,†], Cheng Zheng[1,†], Cuifang Kuang[1,*], Renjie Zhou[2,3], Mohammad M Kabir[4], Kimani C. Toussaint Jr.[5], Wensheng Wang[1], Liang Xu[1], Haifeng Li[1], Peng Xiu[1], and Xu Liu[1,*]

[1]*State Key Laboratory of Modern Optical Instrumentation, College of Optical Science and Engineering, Zhejiang University, Hangzhou, Zhejiang 310027, China*

[2]*Laser Biomedical Research Center, G. R. Harrison Spectroscopy Laboratory, Massachusetts Institute of Technology, Cambridge, Massachusetts 02139, USA*

[3]*Department of Biomedical Engineering, The Chinese University of Hong Kong, Shatin, New Territories, Hong Kong, China*

[4]*Department of Electrical and Computer Engineering, University of Illinois Urbana-Champaign, Urbana, Illinois 61801, USA*

[5]*Department of Mechanical Science and Engineering, University of Illinois Urbana-Champaign, Urbana, Illinois 61801, USA*

*\*Correspondence and requests for materials should be addressed to C.K. (e-mail: cfkuang@zju.edu.cn) and X.L. (email: liuxu@zju.edu.cn).*

*†These authors contributed equally to this work.*

**Abstract**

Here we report nonlinear focal modulation microscopy (NFOMM) to achieve super-resolution imaging. Abandoning the previous persistence on minimizing the size of Gaussian emission pattern by directly narrowing (e.g. Minimizing the detection pinhole in Airyscan, Zeiss) or by indirectly peeling its outer profiles (e.g., Depleting the outer emission region in STED, stimulated emission microscopy) in pointwise scanning scenarios, we stick to a more general basis------ maximizing the system's frequency shifting ability. In NFOMM, we implement a nonlinear focal modulation by applying phase modulations with high-intensity illumination, thereby extending the effective spatial-frequency bandwidth of the imaging system for reconstructing super-resolved images. NFOMM employs a spatial light modulator (SLM) for assisting pattern-modulated pointwise scanning, making the system single beam path while achieving a transverse resolution of ~ 60 nm (~$\lambda$/10) on imaging fluorescent nanoparticles. While exploring a relatively simple and flexible system, the imaging performance of NFOMM is comparable with STED as evidenced in imaging nuclear pore complexes, demonstrating NFOMM is a suitable observation tool for fundamental studies in biology. Since NFOMM is implemented as an add-on module to an existing laser scanning microscope and easy to be aligned, we anticipate it will be adopted rapidly by the biological community.

**Text**

In past two decades, a number of super-resolution methods have been demonstrated through manipulating the illumination beams, and thus modulating the emission patterns [1-5] in continuum-regime (where the sample is treated as a continuous object in the recording) [6]. Continuum- regime illumination patterns can be categorized into structured illumination (SI) and point illumination (PI), as widely implemented in structured illumination microscopy (SIM) [7, 8] and confocal microscopy [9], respectively. SI can be considered as a simplified and optimized type of PI that collects all possible spatial frequencies within the cut-off band by projecting illumination patterns of different structures [refer to Supplementary Material (SM), Sec. 1]. A high illumination density is crucial to achieve nonlinear excitation in order to surpass a 2-fold resolution enhancement as explored in both stimulated emission depletion (STED) [1] microscopy and saturated structured illumination (SSIM) [10, 11] microscopy. Evidenced by the developments of STED, PI is a better choice to implement fluorescent saturation techniques, since it can easily offer a high illumination density through a strongly focused illumination. But STED is not flexible on fluorescent dyes with certain laser sources and SSIM needs a large number of recordings while using a high illumination power, and both methods deploy complex systems, thus limiting them from a broader adaption. Recently, several groups have implemented SIM-based algorithms for post-processing in imaging scanning microscopy (ISM) [12-14] (a modification of confocal) to obtain extraordinary efficacy of resolution enhancement under linear [15, 16] and nonlinear [17] excitation processes. However, in their systems the use of a Gaussian illumination pattern has resulted in an emission pattern with weak high-frequency components, which prohibits substantial improvement of resolution under fluorescence saturation condition.

Here we overcome the previous issues by finding a feasible and flexible scheme. Namely, we ask the question – given a focused PI scenario with certain characteristics--which focal emission scheme would yield more physical information of an object while being simple to be implemented? When retrospecting previous PI super-resolution methods, we could conclude they were all keen on achieving one thing, that is minimizing the size of emission point spread function (PSF) with every possible means: directly sharpening by Bessel beam [18], super-oscillating beam [19, 20], polarization modulation [21], minimizing the size of pinhole, imaging through the scattering media [22-24] or peeling the outer profiles of the emission PSF (STED and so on). However, when considering the frequency shifting framework, it turns out that it isn't necessary to anticipate the focused emission PSF to be slender enough to tell us about resolution as previous methods do, but instead, maximizing the overall coverage of emission patterns in Fourier domain should be a more comprehensive guideline [6].

Thus, in this letter we emphasize the frequency shifting as a general basis in the PI microscopy and introduce nonlinear focal modulation microscopy (NFOMM), a technique primarily based on using high-intensity modulated-pattern scanning. In NFOMM, through proper focal light-field modulations, prominent high-frequency components and a large cut-off bandwidth have been realized under the nonlinear illumination-excitation-emission process. Similar to SSIM, NFOMM shifts the sample's high spatial resolution information into the detection passband. Finally, we post-process acquired images with a forward model described by the $OTF_{eff}$ to reconstruct the object.

*Principle.*-- In a classical PI microscope without a specific illumination modulation, the sample fluorescence is excited by a focused Gaussian beam (Fig. 1a,iv) with a zero-phase mask [Fig. 1(a,i)] and the fluorescence emission is detected at the image plane through a microscope objective. Another two illumination patterns can be generated by modulating the phase of incident beam with the phase masks as shown in Fig. 1(a, ii-iii). The PSFs associated with these patterns are simulated within vectorial diffraction theory [25, 26], and the parameters for the fluorescence saturation characteristics are from Ref. [27]. When increasing the power of the two phase modulated illumination patterns, the emission patterns from the sample are nonlinearly modulated by both the doughnut/line shape illumination and the fluorescence saturation. The corresponding effective system optical transfer functions ($OTF_{eff}$) of the above three illumination patterns are shown in the third row of Fig. 1(a), where we can find that the doughnut outputs stronger high-frequency components and the y-direction line-shape PSF covers more high-frequency components than that of the other two illumination patterns along the x-direction. Fig. 1(b) shows the normalized x-direction line profiles of the $OTF_{eff}$ to intuitively compare the performance between three different illumination patterns. When the illumination intensity is increased to $100\,\text{kW/cm}^2$, the nonlinear effect due to fluorescence saturation extends the $OTF_{eff}$ of Gaussian emission to be beyond that of the non-saturated Gaussian emission. The saturated doughnut emission pattern as well as the saturated line-shape emission pattern expands the frequency band, but importantly their high-frequency components increase more than the saturated Gaussian emission pattern. Therefore, the two saturated emission patterns give a wider $OTF_{eff}$ distribution, leading to better resolution. Notably, the lack of high-frequency components in

$OTF_{eff}$ with saturated Gaussian illumination (GI) pattern explains why the resolution enhancement of saturated scanning SIM is limited to 2.6 folds [17]. This also explains why the resolution of saturated GI based saturated excitation microscopy (SAX) has not yet achieved a resolution beyond 140 nm at an 532 nm excitation wavelength [28].

The high-intensity doughnut/line illumination patterns induce nonlinear effects that shift the sample's high-frequency components into the passband of the $OTF_{det}$. Next, an image post-processing algorithm is essential for deconvolving the raw data to obtain the final super-resolved images. The reconstruction can be considered as an inverse problem. *i.e.*, an estimation of the original sample structure according to the system's forward model and the detector recordings. To efficiently obtain information from the recordings, we apply a positively constrained Richardson-Lucy (RL) deconvolution algorithm with the forward model (see details of the RL reconstruction and the forward model in SM Sec. 3).

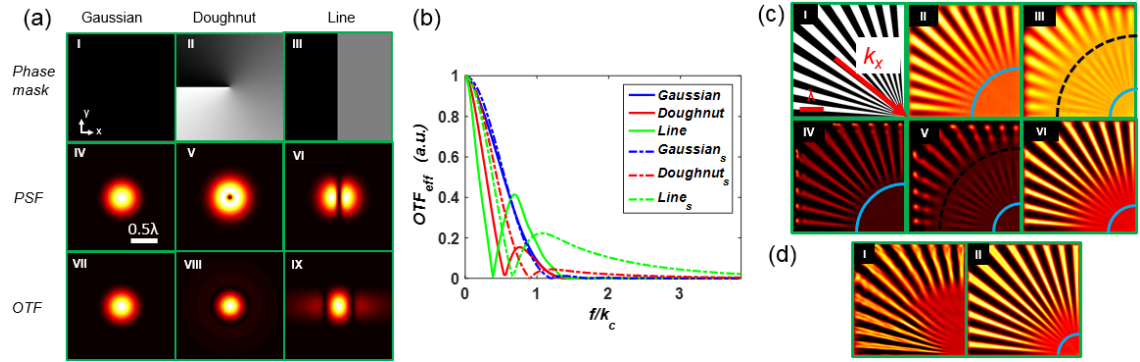

FIG. 1. *Working principle*. (a) (i-iii) are zero, 0-2π vortex and 0-π step phase masks and (iv-vi) and (vii-ix) are the resulting $PSF_{eff}$ and $OTF_{eff}$. The illumination peak intensity of Gaussian PSF is 3 kW/cm$^2$ while that for Doughnut and Line shape PSFs is 100 kW/cm$^2$. Size of detecting pinhole is assumed to be 0.6 AU-diameter. (b) Normalized profiles of the $OTF_{eff}$ corresponding to non-saturated Gaussian (blue line), non-saturated doughnut (red line), non-saturated line-shape (green line) illumination patterns at the illumination peak intensity of 3 kW/cm$^2$ and their counterparts (dashed lines) under saturated condition (100 kW/cm$^2$). (c) Upper row, imaging results of the spoke-like sample [indicated in (c,i)] with GI at a peak intensity of 3 kW /cm$^2$ [indicated in (c,ii)] and SDI at a peak intensity of 100 kW/cm$^2$ [indicated in (c,iii)], respectively; lower row, (c,iv-v) are recovered RL deconvolution results of (c,ii-iii), respectively, (c,vi) is the NFOMM result generated from (c,ii-iii) with multi-view deconvolution. (d,i) the recovered result of the SLI image at an illumination peak intensity of 100 kW/cm$^2$ with the RL deconvolution; (d,ii) the NFOMM result reconstructed by the multi-view deconvolution, using recordings generated by the SLI patterns with four orientations and a peak intensity of 100 kW/cm$^2$, a GI illumination pattern with a peak intensity of 3 kW/cm$^2$. Here, $k_x$ in (c,i) denotes the direction of the object spatial-frequency increase. The light blue circles (c, d) denote the borders beyond which one can barely discern object details, and the related frequency component will be cut off. The black dashed circles (c) denote the frequency deficiency regions. The iterations (c, d) are all 200.

A spoke-like sample, shown in Fig. 1(c,i), is used to examine NFOMM's frequency retaining ability. We first make a comparison between the performance of GI and the saturable doughnut illumination (SDI) imaging results. Comparing the images in Fig. 1(c,ii), and Fig. 1(c,iii), we find that the circle diameter in the latter is much smaller than that in the former. More inner details are discernible in Fig. 1(c,iii), which indicates a higher resolution can be achieved. However, the dual crests feature of the doughnut emission PSF induces inner-region malposition in Fig. 1(c,ii), reflecting a contrast inversion in the complex amplitude. In Fig. 1(c,iv) and Fig. 1(c,v), the deconvolution makes the already recognized regions more distinct and the reconstructed spokes thinner, which validates the frequency-lifting ability of the RL deconvolution method [29]. Apparently, NFOMM achieves subtler details without the mal-positioning issue, confirming the superiority of this nonlinear modulation method.

The NFOMM technique has the flexibility to incorporate multiple recordings acquired under differed modulation conditions, which is capable of rendering a superior result. Noting that in NFOMM the retrieved sample frequency with a single SDI recording suffers deficiency at a certain point (corresponding to the valley of the dashed red line in Fig. 1(b), where the frequency strength is orders of magnitude lower than its neighborhoods. Indicated by the black dashed circles in Fig. 1(c,iii and v), the frequency deficiency from SDI contributes to the relatively blurred regions. The deficiency does not severely influence the final image reconstruction result in practical experiments with this SDI-based NFOMM, and it can be also recovered through superposing the Fourier spectrums of the GI recording with that of SDI recording to obtain a synthetized Fourier spectrum. This method is named multi-view NFOMM, *i.e.*, the GI and SDI recordings are regarded as two views. Here we adapt the RL deconvolution into a multi-view deconvolution algorithm, see details about the reconstruction procedure in SM, Sec. 4. Fig. 1(c,vi) shows an instance of this dual-view NFOMM result, where the once annular blurred region has now been compensated.

To further verify NFOMM's super-resolution capacity, we examine the performance of NFOMM with the saturable line-shape illumination (SLI) modulation [(Fig. 1(d)]. As seen in Fig. 1(d,i), the recovered result [Fig. 1(d,i)] from a single $y$-direction SLI recording [Fig. 1(a,vi)] achieves a better resolution along the $x$-direction due to the abundant high-frequency components in this direction. Due to the lack of $y$-direction frequency components, the spokes are blurred and somewhat distorted along the clockwise direction. Therefore, we make four SLI recordings with four illumination orientations (0°, 45°, 90°, and 135° relative to the $x$-direction) and one GI recording for the multi-view NFOMM reconstruction. Shown in Fig.

1(d,ii), the corresponding SLI based multi-view NFOMM achieves improved resolution in all directions with no blurring. Considering that the SLI-NFOMM demands multiple recordings and is more vulnerable to photobleaching and aberrations, this letter mainly evaluates the NFOMM experimental performance with the SDI modulation mode.

*Examining the resolution of NFOMM with the single SDI modulation.* We conduct a series of experiments utilizing our NFOMM system detailed in SM, Sec. 2. First, we image fluorescent nanoparticles [F8789-FluoSphere Carboxylate-Modified Microspheres, 0.04 µm, dark red (660,680)] with a continuous wave source at an illumination wavelength of 635 nm with a pinhole size of 0.74 AU. The NFOMM results [Fig. 2(b)] show a significant resolution improvement compared to the confocal results [Fig. 2(a)]. A large number of nanoparticles that were conglomerated in the confocal images are now clearly resolved. In Fig. 2(c,d), the magnified views of beads in the white boxes provide a more distinct comparison. Also, density profiles along the green dashed lines in Fig. 2(c,d) show a discernible distance of 93 nm (~λ/7) between the two adjacent nanoparticles with an illumination power of 90 µW. These nanoparticles that once appeared as one (in the red confocal curve) now appear as two separated peaks (in the green NFOMM curve) in Fig. 2(e). See more details about Fig. 2 in SM, Fig. 5, where we also include a comparison between NFOMM and deconvolved Confocal.

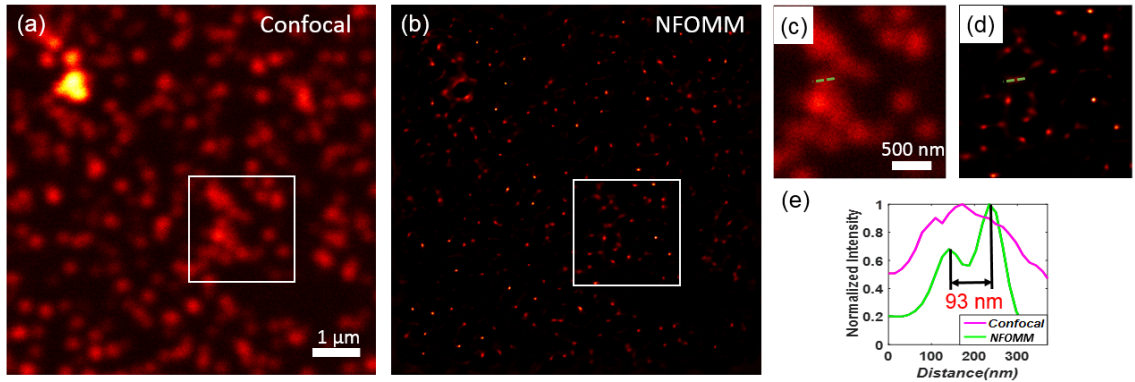

FIG. 2. *NFOMM applied to 40 nm nanoparticles achieves a transverse resolution of λ/7.* (a-b) Imaging results of 40 nm fluorescent nanoparticles. (c–d) Magnified views of areas indicated by white boxes in (a-b), respectively. (e) Density profiles along the green dashed lines in (c-d).

Since the obtained raw signal measured at the detector is directly the saturated emission fluorescence from the stained sample, the SNR of NFOMM is 4-10 folds larger than that of confocal. However, a large fluorescence intensity can potentially induce detector saturation, especially when imaging dense tissues regions. This concern restricts a further improvement of illumination power to achieve higher resolution. To avoid this issue in the experiments, a tunable attenuator was positioned before the detector to lower the emission fluorescence intensity. In Fig. 3, we implement a narrowed pinhole scheme with the pinhole's diameter (0.37 AU) half of that used in the former experiments to alleviate the detection saturation issues at high illumination power. According to the confocal principle [30], a narrower pinhole not only helps with rejecting the out-of-plane fluorescence signal arriving at the detector for lifting contrast, but also directly improves the resolution simultaneously. We apply NFOMM along with the narrowed pinhole scheme for experimental testing of both sparse [Fig. 3(a-e)] and dense [Fig. 3(f-h)] fluorescent samples. As expected, we observe a significant improvement of transverse resolution up to ~ 60 nm (less than λ/10) at an illumination power of 2.1 mW when imaging 40 nm fluorescent beads [Fig. 3(e,h)].

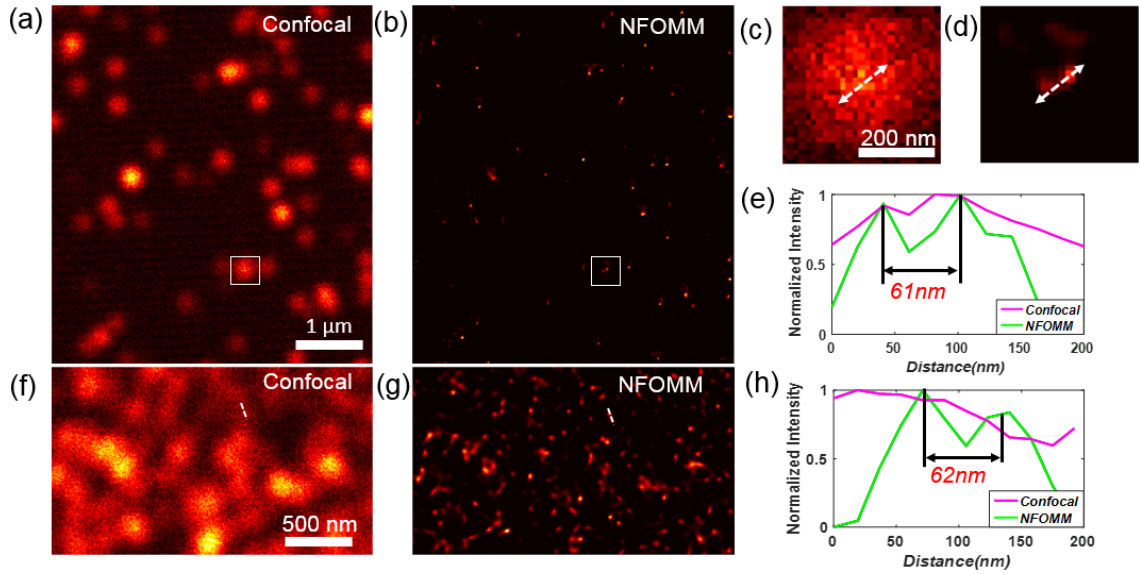

FIG. 3. *NFOMM applied to 40 nm nanoparticles with pinhole size narrowed to 0.37AU achieves a transverse resolution of λ/10.* (a-b) Imaging results of 40 nm fluorescent nanoparticles. (c–d) Magnified views of areas indicated by white boxes in (a-b), respectively. (e) Density profiles along the white dashed lines in (c-d). (f-g) Imaging results upon dense nanoparticles. (h) Density profiles along the white dashed lines in (f-g).

*Biological imaging of NFOMM with dual view SDI modulation.* Next, experiments using Vero cells are performed with dual view NFOMM, *i.e.*, reconstructed the object from the SDI and GI recordings. As expected, imaging results demonstrate that the sub-diffraction resolution in both transverse directions are achieved in biological samples. Fig. 4(a) shows a comparison of resolution enhancement across a whole cell network, where NFOMM reveals the fine structures of the Tubulin networks that are not available in the confocal images. Moreover, a two-color sample is imaged [Fig. 4(b)], where the resolution improvement is well observed in the comparison. The fine structures of both tubulin microtubules (Alexa594 colored) and vimentin (Star635P colored) are more discernable in the later NFOMM images, confirming the capacity of NFOMM in multi-color biological imaging (see more details about the dual color imaging figure in SM, Fig. 7).

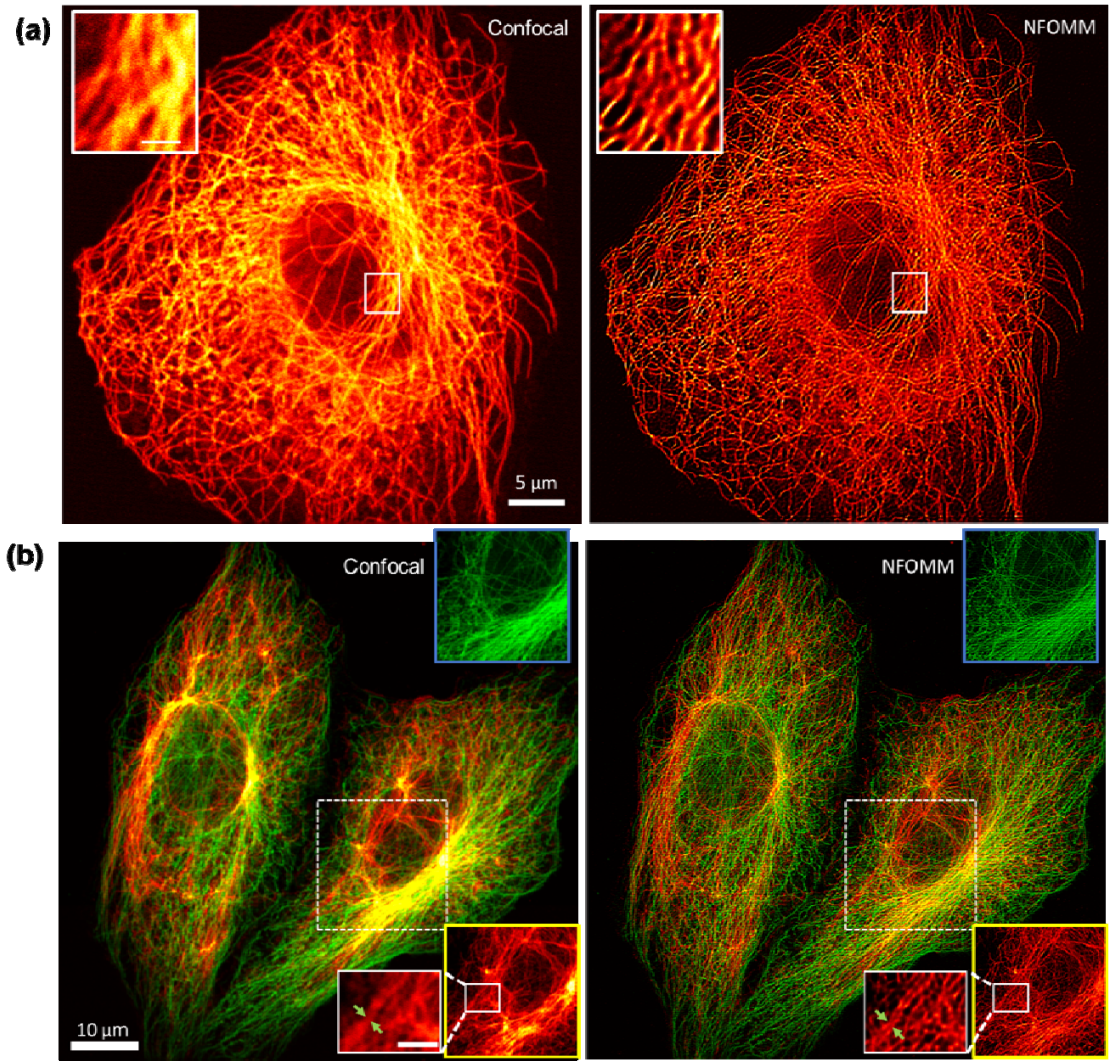

FIG. 4. *Demonstration of NFOMM applied to Vero cells in comparison to using standard confocal microscopy.* (a) Images of tubulin microtubules in vero cells immune-labeled by Star 635P. On the top left of figures in (a) are the magnified views of regions indicated by the white boxes. (b) Dual color imaging results upon samples Alexa 594 immuno-labeled tubulin (green color) & Star 635P immune-labeled vimentin (red color). On the top-left of figures in (b) are results of the tubulins in the region highlighted by the dashed white box, while on the lower left of each figure are results of vimentins of the same region. Inset scale bars represent 1 μm.

*SLI-NFOMM with the blind multi-view reconstruction.*-- Finally, for a proof-of-concept, we demonstrate the resolution enhancing ability with the SLI modulation. As can be seen from the Fourier transforms of the recordings corresponding to different SLI orientations at the saturation illumination power [Fig. 5(a-d)], the saturation effect, along with the phase modulation, successfully extend the imaging frequency border beyond that of the GI [Fig. 5(f)] as well as the SDI [Fig. 5(e)]. Since the previous model based multi-view reconstruction cannot be easily applied onto the SLI recordings due to the aberrations, we implement the blind gradient descent (BGD) based multi-view reconstruction algorithm here (see details about the BGD algorithm in SM, Sec. 5). When comparing Fig. 5(h,i) with the Fig. 5(g), we observe a significant increase in resolution with the NFOMM result. Highlighted in the white/green boxes in Fig. g-i, the regions that once blurred in the confocal image are well discerned in the accompanied SDI-NFOMM [Fig. 5(h)] and SLI-NFOMM [Fig. 5(i)] results. Furthermore, the resolving superiority of the SLI-NFOMM to that of SDI-NFOMM [see the green boxes in Fig. 5(h,i)] well resonates with the previous reasoning in the principle part. This provides a superior way to generate patterns for NFOMM in future study.

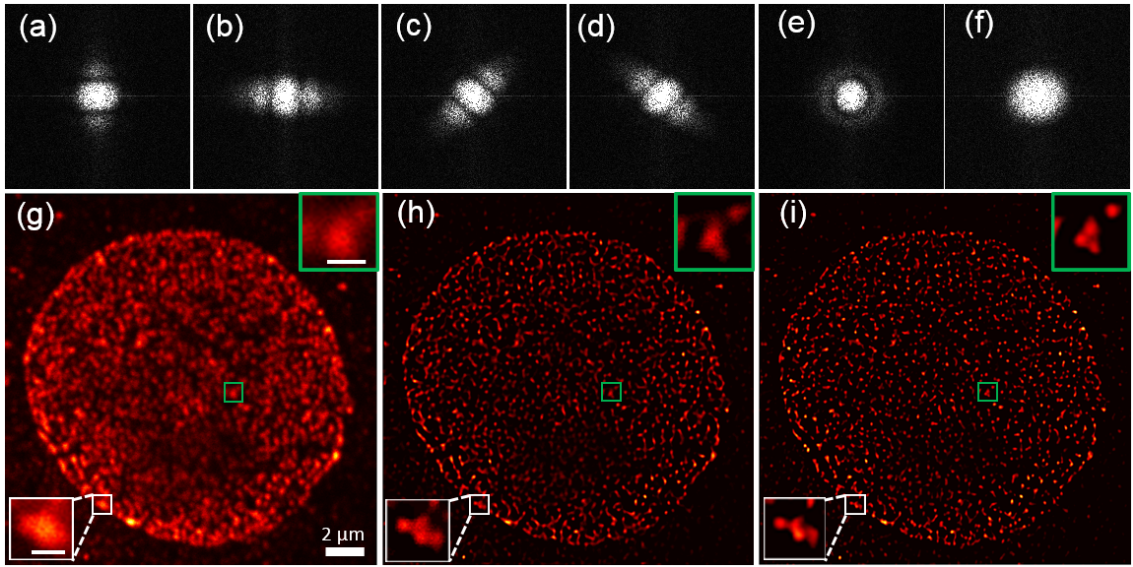

FIG. 5. *Experimental examination of the SLI-based NFOMM.* (a-e) Fourier spectrums of the recordings of nuclear pore complexes in vero cells immune-labeled by Star 635P with 0° direction SLI, 90° direction SLI, 135° direction SLI, 45° direction SLI and the SDI. (f) Fourier transforms of the recorded images of vero cells with the GI. (g) Confocal imaging results corresponding to (f). (h) NFOMM result using the BGD algorithm and the recordings corresponding to (e), (f). (i) NFOMM result using the BGD algorithm and the recordings corresponding to (a-d, f). Inset scale bars in g are 500 nm.

*Discussion.--* In this very first demonstration, NFOMM has already achieved resolution exceeding that of SIM, ISM [31], SAX [32]. To validate NFOMM's performance, we have compared the imaging results with the pulsed wave STED method using the same sample (SM Fig. 7). As expected, NFOMM results have achieved a comparable resolution to STED. In continuous-wave STED [33], the FWHM of 60 nm was enabled at a depletion power of 160 mW according to the square root law [34], while in NFOMM we have achieved a similar resolution with a power of only 2.1 mW (Fig. 3). Despite a different experimental environment, e.g., temperature, medium and sample variations, the lower power requirement can be attributed partially to a larger fluorescence excitation absorption cross section in NFOMM compared to the depletion absorption cross section in STED [33]. Unlike the depletion [1] or competition methods [5] that are keen on sharpening the Gaussian emission pattern with a saturated depletion beam, NFOMM possesses a considerably high SNR due to detecting the lossless saturated fluorescence that emits from the sample. However, in the experiments, the NFOMM may risk lowering the sectional ability due to the saturation effect, which can be eliminated by shearing the Fourier spectrums of the lower frequency region of GI recording and the higher part of the modulated recordings (see more details in SM Sec. 6). Moreover, in SM Fig. 9 we give an evaluation of photobleaching on imaging vimentins as a reference.

To verify NFOMM's applicability, we applying NFOMM on imaging 3D biological structures of the Vero tubulin networks (SM Fig. 10, SM Video 1) and the Vero nuclear pore complexes (SM Video 2) with lateral modulation. NFOMM achieves both high contrast and high transverse super-resolution owing to the frequency shifting and the deconvolution procedures. Moreover, the 3D super-resolution capacity of NFOMM is preliminarily verified with simulations (SM Sec. 7). Our future work aims to experimentally realize 3D super-resolved NFOMM and its applications in biological observations. The proposed NFOMM is compatible with point-scanning based methods such as ISM or Rescan [35], which may also be able to alleviate the frequency deficiency issues (see simulations in SM, Fig. 11). Moreover, owing to its single laser source and less-constraint on the perfectness on the illumination pattern, NFOMM is generic and applicable to two-photon, light-sheet, and other compatible methods imaging saturable media (e.g., Graphene, Perovskite and even gold scattering) for circumventing the diffraction limit.

In summary, we have developed NFOMM, as a computational super-resolved imaging method, to achieve $\lambda/10$ resolution. Given its simplicity, low illumination power, and promise to be easily added onto existing laser scanning microscopes, we envision that NFOMM will be quickly adapted and greatly facilitate biological/material observations for fundamental studies in the future.

**Acknowledgements**
G. Z. appreciates Dr. Rainer Heintzmann (Friedrich-Schiller-Universität Jena) for his discussion about SIM-like post-processing algorithms with data obtained by pointwise scanning, and Dr. Jan Keller-Findeisen (Max Plank Institute For Biophysical Chemistry) for his discussion about requirements of fluorescent dyes in STED&SSIM. We also thank Abberior Instruments for preparing the samples and the STED experiments. This work was financially sponsored by the Natural Science Foundation of Zhejiang Province LR16F050001, the National Basic Research Program of China (973 Program) (2015CB352003), National Key R&D Program of China (2017YFC0110303, 2016YFF0101400), the National Natural Science Foundation of China (61378051, 61427818, and 61335003), the Fundamental Research Funds for the Central Universities (2017FZA5004), and the US National Institute of Health (NIH) grants NIH9P41EB015871-26A1.